\documentclass[conference]{IEEEtran}

\usepackage{multicol}

\usepackage{xcolor}
\usepackage{soul}
\usepackage{amsmath}
\usepackage{graphics}
\usepackage{setspace}
\usepackage{cite}
\usepackage{latexsym}
\usepackage{float}
\usepackage{epsfig}
\usepackage{multirow}
\usepackage{cite,cases,url}
\usepackage{amssymb}
\usepackage{graphicx}
\usepackage{epstopdf}
\usepackage{balance}
\usepackage{enumerate}
\usepackage{array}
\usepackage{algorithmic}
\usepackage{algorithm}
\usepackage{gensymb}
\usepackage{siunitx}
\usepackage[normalem]{ulem}
\useunder{\uline}{\ul}{}

\newcolumntype{L}{>{\centering\arraybackslash}m{5cm}}
\newcolumntype{K}{>{\centering\arraybackslash}m{6cm}}
\newcolumntype{P}{>{\centering\arraybackslash}m{2.3cm}}
\newcolumntype{M}{>{\raggedright\arraybackslash}m{2cm}}
\newcolumntype{N}{>{\raggedright\arraybackslash}m{2.5cm}}

\begin{document}

\title{Machine Learning-Assisted UAV 
Operations with UTM: 
Requirements, Challenges, and Solutions}

\author{\IEEEauthorblockN{Aly Sabri Abdalla,
Vuk Marojevic\\
}
\IEEEauthorblockA{Department of Electrical and Computer Engineering, Mississippi State University, MS 39762, USA.\\}
Email: \{asa298$|$vm602\}@msstate.edu
}

\maketitle


\begin{abstract}
Unmanned aerial vehicles (UAVs) are emerging in commercial spaces and will support many applications and services, such as smart agriculture, dynamic network deployment, and network coverage extension, surveillance and security. 
The unmanned aircraft system (UAS) traffic management (UTM) provides a framework for safe UAV operation integrating UAV controllers and central data bases via a communications network. 
This paper discusses the challenges and opportunities for machine learning (ML) for effectively providing critical UTM services. 
We introduce the four pillars of UTM---operation planning, situational awareness, status and advisors and security---and discuss the main services,  specific opportunities for ML and the ongoing research. 
We conclude that the multi-faceted operating environment and operational parameters will benefit from collected data and data-driven algorithms, as well as online learning to face new UAV operation situations. 
\textit{Index Words}--
UAS, UAV, UTM, ML.
\end{abstract}

\IEEEpeerreviewmaketitle

\section{Introduction}
\label{sec:intro}

Over the past decade there have been extensive contributions into the air transportation system that directly benefits passengers and businesses who rely on aviation every day. However, most of airspace merging endeavor have emphasized on integrating medium or huge unmanned aircraft into non-segregated civil airspace (above 400 feet) where most of civil and military aviation activities happen\cite{FAA_Airspace}. On the other hand, it is expected that within the next five to ten years we will enter the era of Unmanned Aerial Vehicles (UAV), and Unmanned Aircraft System (UAS) \cite{jaber2017}. 

The current and expected applications of UAVs include delivering cargoes and food to customers' doorsteps, monitoring and surveillance missions throughout a city, and enhancing the cellular and Internet connectivity to ground users. The majority of UAV applications will occur below 400 feet in the segregated civil airspace\cite{UAVCivil}. Therefore, 
collaborations between the Federal Aviation Administration (FAA), National Aeronautics and Space Administration (NASA), other federal partner agencies, and industry  are ongoing to investigate and create the Concept of Operations (ConOps), information exchange requirements, and a framework to enable beyond visual line of sight (BVLOS) UAV operations below 400 feet. 

To this end, the UAS Traffic Management (UTM) ecosystem has been developed and upgraded to formally define functions, responsibilities and roles, data exchange architecture and protocols, and requirements needed for managing the uncontrolled UAVs operations at low elevation\cite{NASA_UTM}. The UTM functionality is built on various levels of information sharing to ensure safe operations; for instance, data exchange from one UAV controller to another, from one UAV to another, from a UAV controller to the FAA. 
The UTM operations are vital component in the UAS, specifically for the BVLOS operations where the operators will not be able to 
distinguish between UAV and manned aircraft. Therefore, UAS BVLOS operations need to rely on UTM services. Those services include, but are not limited to registration, airspace authorization, identification of other nearby UAS operators for data exchange, Remote Identification (RID) transmission, strategic deconfliction through the sharing of flight intent and negotiation, monitoring of conformance to flight intent, notification/alerts of in-flight conflicts, in-flight reroute, weather, surveillance, and navigation\cite{FAA_Con}. Some of these services can be completed before UAV launch; 
however, BVLOS operation will required real-time adjustments and adaptive services. 

Considering the foreseen increment in commercial UAS flights and tasks, specifically at low heights in the airspace \cite{jaber2017}, the roles and responsibilities of the UTM becomes more critical. 
The UTM functions need to provide reliable, secure, and large-scale services to maintain safe and efficient operations. Therefore, in order to guarantee a high-performing UTM framework that is scalabile, of low overhead and energy consumption, and capable of real-time processing of shared data, advanced technologies need to be integrated with the UTM. Machine learning (ML) has been widely recognized as a promising solution for supporting potentially transformative advances in different areas, such as healthcare, autonomous vehicles, wireless communications, and image and speech recognition. 
ML techniques are effective for complex management task, where accurate models are lacking, and where training data is available or can be obtained.
\begin{figure*}[h]
    \centering
    \includegraphics[width=0.6\textwidth]{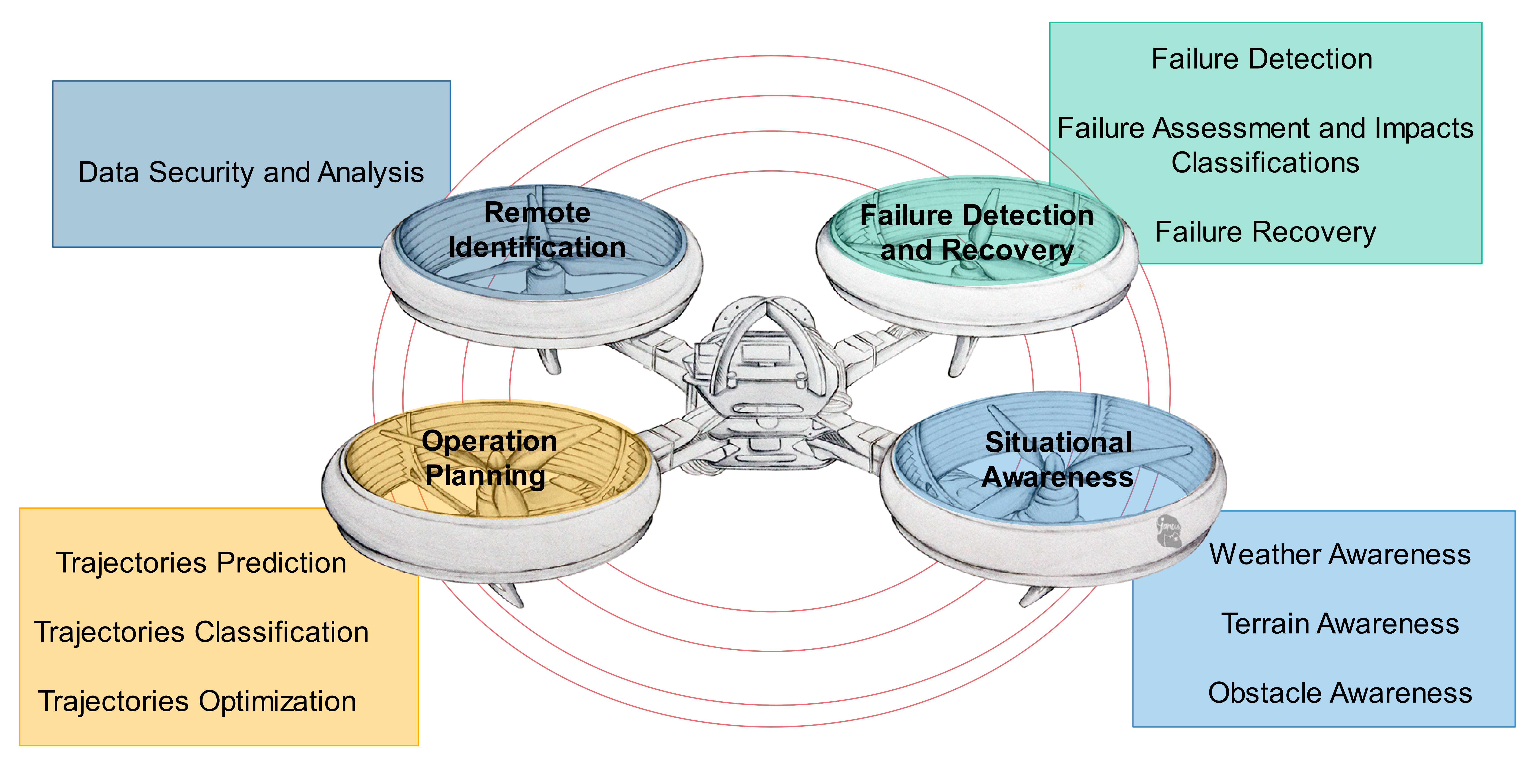}
    \vspace{-5mm}
    \caption{
    UTM functions and services.}
    \vspace{-5mm}
    \label{fig:Figure1}
\end{figure*}

In this paper, we investigate the requirements of real-time interconnection between BVLOS UAVs operations and the UTM for enhancing the safety of operations. Also, we provide an initial survey of how machine learning techniques can be used to boost the security, reliability, and accuracy of UAV-UTM interconnections.


\section{Machine Learning-Assisted Framework}
\label{sec:system}

In this section, we discuss each of the available UTM functions and discuss ML algorithms that can be used for enhancing the overall performance of the system. 
Fig. \ref{fig:Figure1} depicts the established UTM functions and services and Table~\ref{tab:my-table} summarizes the ML opportunities and techniques to support these services.

\subsection{Operation Planning}
The operation plan which is also known as"Flight plan" is one of requirements that must be submitted to the UTM entity prior to the operation. The operation plan is formulated as a group of four-dimensional (4D) segment of the airspace where the UAV will execute its operation. The UAV's trajectory flight segment will include the three component of each location and the corresponding time when the UAV will be at each position. The flight plan of each UAV operation will include information related to the launch and recovery of the UAV. The operation plan will be shared with other entities to ensure the efficient use of the airspace and minimize the possibilities of intersect with other aircraft operations. 
The operation plan can be effected by several factors, for instance, other aircraft operations, airspace constrains which may include some limitations for special use of the airspace, land restrictions such as social assemblages and restricted flying areas. First, those types of information will be collected and reported to the UTM for adjusting the operation plan. However, processing this amount of information prior to the operation may not be adequate for fully satisfying all safety concerns. 

Therefore, applying ML algorithms can be highly supportive and efficient in such cases where there are many factors and enormous amount of information involved in action planning\cite{MLBigData}. In addition to that, the use of ML algorithms for the UAV's operation planning is vital, specifically in the cases where collecting or getting the operation trajectory information is challenging or not applicable , such as 
operations in hostile environment\cite{MLHostile}. Also, it can assure reduction in the computational complexity required for efficient search of vigorous feasible paths.

\begin{table*}[ht]
\centering
\caption{ML opportunities and techniques for the UTM functions of Fig. 1.}
\vspace{-2mm}
\label{tab:my-table}
{\begin{tabular}{|p{2cm}|p{6.6cm}|p{7.8cm}|}
\hline

\textbf{Functions} & \textbf{ML opportunities} & \textbf{
ML techniques}            
\\ \hline

\vspace{-0.02 in}
Operation Planning  &
\vspace{-0.05 in}

\vspace{-0.05 in}
\begin{list}{\labelitemi}{\leftmargin=0.2em}
    \item {\small Dynamic trajectory planning in new situations}
    \item {\small Lack of prior information of paths, needed to classify}
    \item {\small Adjustment of paths as a function of UAV states}
    \item {\small Flight optimization on the fly}
    \vspace{-0.1 in}
\end{list}    

    &
    %
    \vspace{-0.09 in}
\begin{list}{\labelitemi}{\leftmargin=0.2em}
    \item {\small RL when no a priori data is available or the environment changes~\cite{RLPATHprediction}.}
    \item {\small DRL to processes many  parameters~\cite{PathOptimize}.}
   
    \vspace{-0.1 in}
    \end{list}  
    
    \\ \hline
    
    \vspace{0.001 in}
Situational Awareness  & 
\vspace{-0.05 in}

\vspace{-0.05 in}
\begin{list}{\labelitemi}{\leftmargin=0.2em}
\item {\small Weather condition awareness and prediction}
\item {\small Terrain awareness and assessment of communications and safety implications}
\item {\small Obstacle awareness}
    \vspace{-0.1 in}
\end{list}

&
\vspace{-0.09 in}
\begin{list}{\labelitemi}{\leftmargin=0.2em}
\item {\small CNN to estimate the 3D wind flow for trajectory adjustment~\cite{WindCNN}.}
\item {\small SVM to classify LoS and NLoS link conditions of the UAV along the flight~\cite{TerrianSVM}.}
\item {\small DL, RL, and CNN for obstacle detection and avoidance~\cite{ObstacleDRL}.}
    \vspace{-0.1 in}
\end{list}

\\ \hline

 \vspace{0.001 in}
Failure Detection and Recovery  &
\vspace{-0.05 in}

\vspace{-0.05 in}
\begin{list}{\labelitemi}{\leftmargin=0.2em}
\item {\small Failure detection (UAV, comm. network, ground users)}
\item {\small Failure assessment and impacts classifications}
\item {\small Failure recovery}
    \vspace{-0.1 in}
\end{list}

&
\vspace{-0.09 in}
\begin{list}{\labelitemi}{\leftmargin=0.2em}
\item {\small Supervised learning that uses historical datasets for prediction of future failures~\cite{HWFailure,HWFailure2}.}
\item {\small 
ANN 
for risk assessment in the UAS network~\cite{MLRISK}.}
\item {\small RL 
to define the optimum fault-tolerant recovery policy against different failures or attacks~\cite{FailurReco}.}
    \vspace{-0.1 in}
\end{list}

\\ \hline

Remote Identification   &
\vspace{-0.05 in}

\vspace{-0.05 in}

\begin{list}{\labelitemi}{\leftmargin=0.2em}
\item {\small Data security and analysis}                      \vspace{-0.1 in}
\end{list}
                
& 


\begin{list}{\labelitemi}{\leftmargin=0.2em}
\vspace{-0.09 in}
\item {\small Data mining techniques to provide a secure analytic of dynamic streams of data~\cite{datamining}.}
\vspace{-0.1 in}
\end{list}

\\ \hline
\end{tabular}%
\vspace{-4mm}
}
\end{table*}
The use of machine learning algorithm to assist UTM's operation planning functions can be classified into: Trajectories prediction, trajectories classifications, and trajectories optimization. 
\begin{itemize}
    \item \textbf{Trajectories prediction:} The ML algorithms can be utilized for online or a proactively trajectory segments prediction in the absence of any prior information for the UTM planning. Some of those aerial tasks are wildfire monitoring, target tracking, or search and rescue. For such cases, the trajectory information are typically minimal or unavailable. To overcome this problem, Reinforcement Learning (RL) is one of the three fundamental ML structures. It is used when no prior information exists. The authors of~\cite{RLPATHprediction} propose applying RL to learn and completer some of the high-level complex tasks, specifically the autonomous navigation in unknown environments.
    
    \item \textbf{Trajectories classifications:} Different from path prediction, ML can be adopted for classifying different available paths, expressly when the UAV operation is conducted in areas where information is not shared. In addition to the shortage of environment information that is needed for prior planning, UAVs are often disposed to internal failures of the hardware and software or the on-board unit that is responsible to communicate with the UTM system. In such circumstances, the UTM node should be able to predict proactively any UAV failure before it happens. This will be discussed later.
    After the failure prediction, the UTM must assess all the predicted paths based on current location of the UAV then evaluate the safest path for task completion or the recovery mechanism. The work presented in~\cite{PathClass} studies the use of a deep neural network (DNN) for classifying and selecting the optimum recovery path for a malfunctioning UAV.  
    
    \item \textbf{Trajectories optimization:} This function can be used to improve the UTM operation planning functionality for designing the UAV trajectory, whether or not prior information exists. The optimization of the UAV's path can be applied for minimizing the flying cost of the UAV which is mainly related to the on-board battery of the UAV. 
    The short battery life of the UAV can be a vital factor that limits the UAV capabilities and availability in the air. This restriction can play a major role specifically in the mission related to search and rescue or disaster management. 
    In addition to that, optimizing the UAV through ML algorithms will consider the authorized airspace restrictions, other aircraft paths, and various other factors that are involved in the planning operation to maintain the safety of the UAV, other aircraft, ground pedestrians and properties, and the task completion. A deep RL (DRL) model, which is the integration of a DNN and RL, is proposed in~\cite{PathOptimize} for reaching the optimum trajectory policy that maximizes the energy efficiency of the UAV. 
\end{itemize}
\subsection{Situational Awareness}
\label{sec:awaremess}
Another important set of UTM functions 
are under the situational awareness category. 
The situational awareness functions are crucial for all of UAS operations. They allow a recognizing any unexpected condition during the UAV operation. Also, it helps determining whether environmental
conditions or other factors are appropriate for conducting the intended operation with regard to the 
location at the specific date and time. 
The situational awareness functions of the UTM rely on shared information from UAVs during operation and from manned aircrafts through UAS Service Suppliers (USSs) \cite{FAA_Con}. 
Such information includes the active path of each aircraft, the intended destinations, and airspace authorizations. 
It is collected during the pre-flight procedure to define the appropriate flight paths.  
However, this does not guarantee a satisfactory foreknowledge of the conditions along the intended flight paths where unexpected difficulties may arise 
that present new challenges during operation. 
Therefore, applying ML will provide a dynamic proactive situational awareness to enhance the safety of operations. 
The situational awareness functions can be classified as: 
\begin{itemize}
\item \textbf{Weather awareness:} Many UAV operations are conducted at low-altitude, where the weather environment can be more complex than at higher altitudes. In addition to that, the small size and weight of small UAVs makes them more vulnerable to 
rapid changes in the weather conditions. 

Weather awareness includes different elements, including dynamic awareness of wind conditions, turbulent weather, and climate zones. 
This information is invaluable to identify the appropriate air spaces and times in advance, before such weather complexities take place. 
The authors of~\cite{WindCNN} estimate the 3D wind flow using a deep Convolutional Neural Network (CNN) in less than two minutes to enable safer trajectories for the UAV during strong wind conditions. 
Reference~\cite{WindReq} highlights that the 
time required for safer UAV flight path determination must be done in less than 30 seconds.  

\item \textbf{Terrain awareness:} The awareness of elevation in the local terrain 
allows to design deconfliction strategies or rerouting paths within the available maneuver space. 
Terrain awareness provides the capability to establish a connectivity map between the vehicle and the UTM along the flight. More precisely, it allows predicting the connectivity holes due to terrain or other constrains, where updates and operational instruction may not be received. 
The loss of communication between the UAV and the UTM during a BVLoS operation can be extremely harmful and result disastrous to the UAV. 
The work presented in~\cite{TerrianSVM} demonstrates the use of Support Vector Machine (SVM)–based ML methods to classify line of signt (LoS) and non-LoS links of the UAV at the different 3D locations and along different 
flight routes. Such classifications can be used to build a 3D connectivity map of the area of operation 
for providing better awareness for the on-board navigation instruments. 

\item \textbf{Obstacle awareness:} The ability to identify unexpected air or ground obstacles that can be static, moving, or dynamically appearing in the flight zone, and may be a crane, power-line Notice to Airmen [NOTAM], or bird activity. 
The awareness of such objects may limit the mobility freedom of UAVs to avoid collision and may result in an online update of the flight trajectory. 
The online obstacle awareness will need a collaboration between the UTM and different sensor nodes (e.g. cameras) along the target trajectory. 
If the path does not satisfy those requirements, then an online autonomous navigation system for the UAV should be applied using on-board cameras. 
The captured data can be used to predict the remaining segments of the trajectory locally, or be sent to the UTM system, capable of running computational complex algorithms.  
The authors of~\cite{ObstacleDRL} survey research discussing the use of Deep Learning (DL) techniques such as RL and CNN and some of the latest DL data sets for real-time testing and validation of obstacle detection and avoidance mechanisms for UAV operations. 


\end{itemize}

\subsection{Failure Detection and Recovery}
The UTM system is responsible for monitoring a UAV during its entire operation. 
Also, the UAV should share with the potentially impacted entities or aircrafts any changes that may effect the safety of the airspace. 
These updates are advertised as Unmanned Aircraft Reports (UREPs)~\cite{UREPS}. Moreover, in case of the occurrence of possible risks on the ground or in the air (police activity, emergency response, and public safety) that would likely influence the safety interests of the UTM or the UAV operation, UAS Volume Reservations (UVRs) will be created 
for notifying those risk in the different airspace volumes.

If any failure occurs in the system, subsystem, or component, the UTM should be able to identify it and report it to the effected entities to avoid massive damage. Those failures can happen at the vehicle or at the UTM itself and include failures of communication with other vehicles or ground stations or UAV hardware failures, such as motor failures or power system failure.  
There are other types of failures that can occur at the UAV, such as sensor failures (GPS, camera, accelerometers, gyros) that can effect the situational awareness and the collected data. Such sensor failures can harm 
the UTM's ability to assess the risk level in a given situation and prevent it from initiating appropriate actions. 

The UTM need be able to detect the occurrence of failures, be able to assess and classify the impact of the failure on the network, and be able to recover from it.

\begin{itemize}
    \item \textbf{Failure detection:} This part is considering the ability to detect various failures in the UTM-connected network, including failures at the UAV, at other ground entities or at the UTM itself. 
    The UTM should be able to provide a quick and accurate reflex to any failure. 
    The prediction of failures allows to develop a timely and effective recovery plan. We suggest a supervised ML model that uses historical data sets to train the model for future predictions of failures. This approach can help to predict sensor failures on the UAV. On the other hand, telemetry or connectivity failures can be predicted using connectivity maps discussed before. 
    To the best of our knowledge, there are no current works that investigate using ML to predict hardware or sensor failures for UAV operations, but ML has been widely used for other applications~\cite{HWFailure,HWFailure2}.    
    
    \item \textbf{Failure assessment and impact classification:} This function will be applied through the UTM to do a risk assessment of the predicted failures. 
    It involves the prediction of multiple scenarios in the case where the recovery mechanism fail. 
    Using this service through the UTM will help forecasting single or combined effects of various failures. 
    The prediction should determine the impact range and hazards and the entities and nodes that should be notified through the USSs. 
    The formal framework for risk assessment is known as the Failure Mode and Effects Analysis (FMEA)~\cite{FEMA}. 
    Applying ML allows to proactively take precautions against complex failures. The work demonstrated in~\cite{MLRISK} gives a survey about different types of ML algorithms that can be used in different engineering risk assessment processes and concludes that the Artificial NNs (ANNs) are the most common model that can be used for such problems. 
     
    \item \textbf{Failure recovery:} The system should be able to recover from any failure as otherwise it can create disastrous situations where lives may be threaten. A failure can occur at the UAV, which can lead to an impact with ground or aerial nodes, or at ground nodes, which provide support service. 
    The previous two functions will be able to detect the failure in the system and assess its effect. 
    Failure recovery should devise the optimal set of actions to minimize the immediate risks and that of future failures. Those actions should be announce to all affected nodes on the ground and in the air to ensure the system safety. 
    There are a variety of actions that can be taken such as immediate termination of the operation or re-routing for maintenance. 
    ML can help reducing the time needed to select the appropriate plan for different use cases based on the predicted damage. RL is used in~\cite{FailurReco} to train a fault-tolerant control policy on a quadcopter's actuator and sensor against failures or attacks. The approach has potential to harden the system against different types of attacks and failures without data sets or prior training.

\end{itemize}
\subsection{Remote Identification}
This section discusses how the UTM can assist in protecting low altitude UAS operation against physical or cyber attacks. 
The UTM will be able to register and authenticate the UAV node or the UAV operator 
through an electronic identification process called Remote Identification (RID)~\cite{RID}. The UTM uses the RID for enhancing the liability and the monitoring capability of the system, specifically to cope with BVLoS operation when the vehicle and its operator are not located in the same area. 
When the system authenticates and registers a new node/operator, the RID information should be shared within the entities that support the operation for the purpose of identification and tracability. 
The UTM supports the RID through the RID message that contains a set of required fields for the system to be able to trace and identify the vehicle/operator: UAS ID, which is a unique identification number, UAS location, and a timestamp. The regulations that define the exact requirements and the required information for a satisfactory RID process are still under development and the information mentioned here is based on latest report and recommendations by the FAA~\cite{ARC}.          
\begin{itemize}

    \item \textbf{Data security and analysis:} While the standardization of the RID is still in progress, the ARC report~\cite{ARC} specifies two approaches for transmitting the RID information: direct broadcast or published and available through the Internet. 
    Both methods are insecure, because anyone within the radio range of the broadcast transmission or with access to the Internet can get the data. 
    ML can enhance the safety of such critical data or limit its availability only to the authorized entities such as public safety or other airspace authorization systems. 
    Also, the expansion of the number of UAVs, operators, and the associated information within a RID message can create sophisticated data sets that should be accessible within a limited time span. Data mining is a popular application of ML that is commonly used in the era of big data. 
    Data mining can be very efficient for data labeling and segmentation, as well as for data analytics through a combination of supervised and unsupervised algorithms. The work proposed in~\cite{datamining} uses data mining techniques to analyze continuous streams of privacy-preserving dynamic data sets.

    
\end{itemize}



. 

    



\vspace{-5mm}
\section{Conclusions}
\label{sec:conclusions}

UAVs are popular and widely available for for civilian and commercial purposes. However, they pose a safety hazard and strict regulations therefore exist. The currently most promising framework for UAV operation is the UTM, which establishes the components and procedures for a safe UAS traffic management. 
This paper reviews the main UTM functions and services and discusses the requirements, challenges and possible solutions for effective service provisioning. We argue that ML will play an important role because of its capability to effectively process huge amounts of data and complex problems. We identify the opportunities and suitable techniques for the different UTM services. In future work we will develop case studies and models, and use data to analyze the proposed solutions and experimentally validate them on suitable research platforms. 




\balance

\bibliographystyle{IEEEtran}
\bibliography{main}


\end{document}